\documentclass[preprint,aps, preprintnumbers, nofootinbib]{revtex4}
\usepackage[dvips]{graphics}

\newcommand{\lsim}[1]{
\setlength{\unitlength}{12pt}
\begin{picture}(1.4,1.)
\put(.7,-0.3){\makebox(0.0,1.)[t]{$<$}}
\put(.7,-0.3){\makebox(0.0,1.)[b]{$\sim$}}
\end{picture}#1}
\newcommand{\gsim}[2]{
\setlength{\unitlength}{12pt}
\begin{picture}(1.4,1.)
\put(.7,-0.3){\makebox(0.0,1.)[t]{$>$}}
\put(.7,-0.3){\makebox(0.0,1.)[b]{$\sim$}}
\end{picture}#2}
\begin{document}
%\preprint{IRB-TH-15/05}
%\preprint{UB-ECM-PF 05/07}
% \draft

\title{Dynamical dark energy with a constant vacuum energy density}

\author{B. Guberina\footnote{guberina@thphys.irb.hr}}
\affiliation{\footnotesize Rudjer Bo\v{s}kovi\'{c} Institute,
         P.O.B. 180, 10002 Zagreb, Croatia}

\author{R. Horvat\footnote{horvat@lei3.irb.hr}}
\affiliation{\footnotesize Rudjer Bo\v{s}kovi\'{c} Institute,
         P.O.B. 180, 10002 Zagreb, Croatia}

\author{H. Nikoli\' c\footnote{hrvoje@thphys.irb.hr}}
\affiliation{\footnotesize Rudjer Bo\v{s}kovi\'{c} Institute,
         P.O.B. 180, 10002 Zagreb, Croatia}

\begin{abstract}

We present a holographic dark-energy model in which the Newton constant
$G_{N}$ scales in such a way as to render the vacuum energy density a true
constant. Nevertheless, the model acts as a dynamical dark-energy model since
the scaling of $G_{N}$ goes at the expense of deviation  of concentration 
of dark-matter particles from its canonical form 
and/or of promotion of their mass to a time-dependent
quantity, thereby making the effective equation of state (EOS) variable and
different from $-1$ at the present epoch. Thus the model has a potential to
naturally underpin 
Dirac's suggestion for explaining the large-number hypothesis, which demands
a dynamical $G_{N}$ along with the creation of matter in the universe. 
We show that with the aid of
observational bounds on the variation of the gravitational coupling, the 
effective-field theory IR cutoff can be strongly restricted, being always
closer to the future event horizon than to the Hubble distance. As for the
observational side, the effective EOS restricted by observation can be made
arbitrary close to $-1$, and therefore the present 
model can be considered as a ``minimal''
dynamical dark-energy scenario. In addition, for nonzero but small curvature
$(|\Omega_{k0}| \lsim 0.003 )$, the model easily accommodates a transition
across the phantom line for redshifts $z \lsim 0.2 $, as mildly favored by
the data. A thermodynamic aspect of the scenario is also discussed.      
\end{abstract}

%\pacs{97.60.Gb, 14.60 Pq, 04.80.Cc}
\newpage

\maketitle

A symmetry principle of gravitational holography \cite{1} 
serves as a window to a 
complete theory of quantum gravity. According to that principle, the
description of a physical system shows equivalence between a theory
having the gravitational field quantized and a theory defined on the
boundary encompassing a system whose dimension is lower by one. The most
rigorous realization of holography is  
the AdS/CFT correspondence in all events
\cite{2}. An important consequence of the holographic principle is that 
various entropy bounds should be manifest in quantum gravity in the
semiclassical limit. All of them state that complete information stored in
a physical system  scales only with the area encompassing the system.

Motivated by a drastic reduction in effective degrees of freedom as
predicted by gravitational holography, Cohen {\it et al.} \cite{3} showed
that the application of the Bekenstein bound to the maximal possible entropy 
\cite{beken} to
effective field theories can even substantially improve the `old'
cosmological constant (CC) problem \cite{4}. By adapting the Bekenstein
bound they showed that, if a certain relationship between the IR and the UV
cutoff was obeyed, the information from quantum gravity 
could be consistently 
encoded in ordinary quantum field theory. 
Such a relationship prevents the formation of black holes within the effective 
field-theoretical treatment, leading to the bound which is far more 
restrictive than that in \cite{beken}.
With the notion that the size of the
region (providing an IR cutoff) is varying in an expanding universe and that
therefore the vacuum energy density $\rho_{\Lambda }$ promotes to a
dynamical quantity, a few years later their considerations  became a core
of a dynamical CC scenario  generically dubbed `holographic dark energy'.
Derived originally for zero-point energies, the bound predicted by  
Cohen {\it et al.} \cite{3} for $\rho_{\Lambda }$ can be rewritten 
in the form 
\begin{equation}
\rho_{\Lambda}(\mu)=\kappa \mu^2 G_N^{-1}(\mu),
\end{equation}
where $\mu $ denotes the IR cutoff and $\kappa $ represents a degree of
saturation  of the bound. Note that $\kappa $ should be of order of unity
if $\rho_{\Lambda }$ is to account for the present energy density in the
universe. Since $\rho_{\Lambda }$ is always dominated by UV modes, 
Eq. (1) is valid  
irrespective of whether the hierarchy between the
UV cutoff  and particle masses is
normal or inverted.

With the appropriate choice for the IR
cutoff, the setup described by (1) was intended to explain the present
acceleration of the universe \cite{5} and possibly to shed some light on 
the still unresolved problem of 
`cosmic coincidence' \cite{6}. Two different sorts of 
generalization of the above setup can be found in the literature: the 
one \cite{7} also promotes the Newton constant $G_N $ to a dynamical
quantity [as already done in (1)], whereas the other \cite{8} 
relies on some peculiar choices for $\mu $. \footnote{The scenario \cite{8}
was primarily intended to unify the early-time inflation with the late-time
acceleration of the universe.}

The successfulness of the `holographic dark-energy' scenario depends
crucially on the choice for the IR cutoff and on the 
question whether
$\rho_{\Lambda }$ describes perfect fluid or not. The generalization with
the running $G_N $ but canonical matter dilution $(\sim a^{-3})$ 
has the advantage that the IR cutoff is univocally fixed by
the continuity relation once the scaling law for $\rho_{\Lambda }$ (or
$G_{N}$) is known
\cite{9, 7}.  
         
For perfect fluids and for ad hoc chosen cutoffs,
such as the Hubble distance or 
the particle horizon distance, one usually cannot obtain the equation of state
$w$ (EOS) characterizing accelerated universes \cite{10}. For interacting
fluids, models with the Hubble distance show considerable improvement
\cite{11}, and for a variable degree of saturation of the
bound predicted by
Cohen {\it et al.} one can even obtain a transition from a decelerated
to an accelerated era \cite{11}. Another ad hoc chosen cutoff in the form of
the future event horizon seems to work much better for both perfect 
(see second reference in \cite{10}) 
and interacting fluids \cite{13}. Still, most models with ad hoc chosen cutoffs 
suffer
from the `cosmic coincidence problem' or have an EOS too far from $-1$ to
comply with the data. For related works, see \cite{14}.    

In the present paper we study 
the implications of a holographic dark-energy
model in which $G_N \sim \mu^2 $ in Eq. (1) so as to make $\rho_{\Lambda }$ a
true constant. The motivation for such a study is threefold. Firstly, the
above model naturally accounts for recent data which converge rapidly 
towards an EOS $w = -1$, at the same time retaining its dynamic character.
The dynamics of the model stems from the fact that the 
variation of $G_N $ goes
at the expense of deviation of  energy density of the cold dark matter (CDM) 
component $\rho_m $
from its canonical form $\sim a^{-3}$. 
Here three possibilities emerge \cite{15, 16}: 
(i) the total 
number of CDM particles in a comoving volume changes while retaining its
proper mass constant, or (ii) the proper mass promotes to a time-dependent
quantity while retaining the total number of CDM particles constant, or (iii)
both the total number  of CDM particles and its proper mass change. Instead
of dealing with ad hoc chosen IR cutoffs, we find that  
in our model the IR cutoff is univocally
fixed by an amount of deviation of $\rho_m $ from its canonical shape.
Secondly, our model can be considered a minimal model which can account
for Dirac's large-number hypothesis \cite{17}. Namely, Dirac himself
suggested a model with a time-varying gravitational constant $G_N $
supplemented with creation of matter in the universe. Thirdly, our model
also accommodates an exciting possibility of having a transition from
$w > -1 $ to $w < -1 $ at redshifts $z\lsim 0.2 $, 
of which there are already indications in recent data
\cite{18}. Finally, we explore thermodynamic consequences of the model, and
find that the generalized second law (GSL) 
of gravitational thermodynamics demands
creation of matter in the universe (as opposed to destruction), in accord
with Dirac's  suggestion.             

Our starting point for implementing  a dynamical $G_N $ and $\rho_{\Lambda }$ in a
cosmological model is the low-energy effective vacuum action
which is  just the Hilbert-Einstein action with a time-dependent CC and
gravitational constants. Especially relevant are found those models in which
the CC- and $G_N $-variation laws were inferred from some underlying physical
theory, such as perturbative particle physics theory with the
Hilbert-Einstein action treated semiclassically \cite{19, 20}, or the 
quantum-gravity approach where nonperturbative solutions were obtained 
within the
``Hilbert-Einstein truncation''\cite{21}, or gravitational holography
\cite{10,7}.
A particularly interesting model
which appeared recently in \cite{23} also discussed Dirac's cosmology and the
large-number hypothesis,  but in a slightly modified Hilbert-Einstein 
action
containing an arbitrary function of the ratio of the CC over the 4-dimensional
scalar curvature. Let us stress that at any rate  the most popular modeling
for the time-dependent gravitational coupling is through a time-varying
scalar, especially the framework of the Brans-Dicke theory\cite{24}. In
the present paper we stick with the 
Hilbert-Einstein action and the $G_N $-variation law obtained 
from gravitation holography as given by (1),
without introducing any geometrical- or quintessence-like scalar fields.      

The generalized equation of continuity in the framework 
consisting of constant $\rho_{\Lambda}$ but variable $G_N $
[i.e. where $G_N $ scales as
$\mu^2 $ in Eq. (1)] is given by
\begin{equation}
\dot{G}_N(\rho_{\Lambda}+\rho_m)+G_N(\dot{\rho_m}+3H\rho_m)=0 .
\end{equation}
Eq. (2) understands that $G_N T_{total}^{\mu \nu }$ and $T_{\Lambda }^{\mu
\nu }$ are conserved separately \footnote{A similar framework for net
creation of energy was studied in the transplanckian approach to inflation
in \cite{25}, and in gravitational holography in \cite{26}}. 
We find
the scaling $G_N(a)$ from Eq. (2), and the  function $a(t)$ 
from the Friedman equation for flat space
\begin{equation}
H^2=\frac{8\pi G_N}{3}(\rho_{\Lambda}+\rho_m), 
\end{equation}
by making a specific {\it ansatz} for the matter energy density
\begin{equation}
\rho_m=\rho_{m0}a^{-3+\epsilon}, 
\end{equation}
where $\epsilon $ is a constant. Although the parametrization (4) is not the
most general one, it certainly covers a large number of interesting cases,
including small deviations of $\rho_m $ from the canonical 
form.\footnote{The same parametrization has been employed in several recent 
papers (\cite{16,lima} and the first reference in \cite{31})  
considering the variable $\rho_{\Lambda}$ but 
constant $G_N$ scenario. In addition, a strong bound on the 
$\epsilon$-parameter was derived in \cite{opher}.} 
As
mentioned above, $\rho_{m0}a^{-3 + \epsilon }$ 
may understand: $m(a) = m_0 $; $n_m =
n_{m0}a^{-3 +\epsilon }$ or $m(a) = m_0a^{\epsilon }$; $n_m = a^{-3}$ or
$m(a) = m_0a^{\delta }$; $n_m = n_{m0}a^{-3 + \epsilon -\delta }$, where $m$
is the mass of the CDM particles, $n_m$ is their concentration, 
and $\delta $ is another constant. In the
following, we always consider the overall change of $\rho_m $ without
paying attention to the particular cases listed above,
although mass-varying particles deserve attention of their own as they may
even shed a new light on the nature of dark energy \cite{27,16}. 

Using (2), (3), and (4), one  obtains explicit expression for 
$G_N(a)$ and a differential equation that determines $a(t)$; 
\begin{eqnarray}
& G_N(a)=G_{N0} \left( \displaystyle\frac{r_0^{-1}+a^{-3+\epsilon}}{r_0^{-1}+1} 
\right)^{\epsilon/(3-\epsilon)} , & \nonumber \\
& \dot{a}=H_0 a \left( 
\displaystyle\frac{r_0^{-1}+a^{-3+\epsilon}}{r_0^{-1}+1} 
\right)^{3/(2(3-\epsilon))} , & 
\end{eqnarray}
where $r_0 $ is the present-day ratio $\rho_{m}/\rho_{\Lambda}$.

Strong
restriction on the $\epsilon $-parameter can be obtained by considering the
time variation of $G_N$. Writing the time variation of $G_N$ as
$\dot{G}_N/G_N = \gamma H $, 
where $\gamma=-\epsilon/(\rho_{\Lambda}/\rho_m +1)$, 
with the aid of the present
observational upper bound $|\gamma |< 0.1$ \cite{28}
(for astrophysical and cosmological constraints, see e.g. \cite{bertolami}), 
we obtain that 
$|\epsilon | \lsim 0.1$. As
another important observational constraint we may take the redshift
$z_{tr}$, at which the deceleration parameter vanishes, and therefore
denotes a transition from deceleration to acceleration. We obtain
$a_{tr}=(r_0/2)^{1/(3-\epsilon)}$. For $r_0=3/7$, this gives
$z_{tr} = 0.64 $ for $\epsilon = -0.1 $ and $z_{tr} = 0.70 $ for $\epsilon
= 0.1 $, to be compared with the 2$\sigma $ constraint, $0.2 \lsim z_{tr}
\lsim 0.72$ \cite{5}. Although we see that both signs  
of the $\epsilon $-parameter almost equally fit in 
observationally, we  show below that only $\epsilon > 0$ survives 
considerations of gravitational thermodynamics. 
       
Using the holographic dark-energy relation (1), Friedman equation (3)
and the ansatz (4), one finds a
compact formula for the IR cutoff 
\begin{equation}
\mu=\sqrt{\frac{3}{8\pi\kappa}} \frac{H}{\sqrt{1+r_0 a^{-3+\epsilon}}} .
\end{equation}
Note the explicit dependence of $\mu $ on $\epsilon $.
%$\mu $-dependence of the $\epsilon $-parameter. 
In Figs. 1 and 2 we depict the dependence of $\mu^{-1} $ on $a$ along with the 
dependence on $a$ of the other two most-popular ad hoc chosen cutoffs,  
%$d_h(a)$ and $d_e (a)$, 
namely,
the future event horizon and the Hubble distance,
for both signs of $\epsilon $. In both cases we find
that our IR cutoff as given by (6) is always closer to the inverse future
horizon than to the Hubble scale $H$.    

\begin{figure}[h]
\centerline{\includegraphics{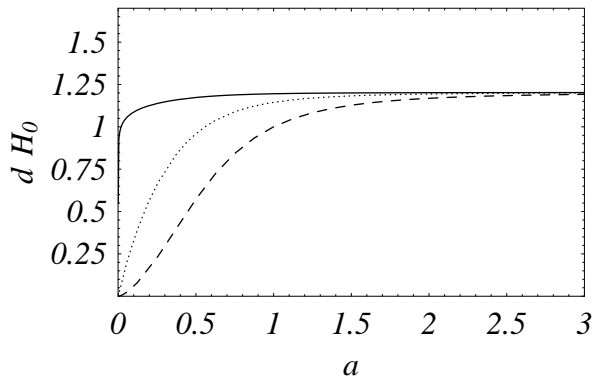}}
\caption{The evolution of various cosmological scales 
$d$ in units of $H_0^{-1}$ as functions of $a$, 
for $r_0=3/7$ and $\epsilon=0.1$.
The future event horizon is represented by the dotted curve,
the scale $\mu^{-1}\sqrt{3/8\pi\kappa}$  
by the solid curve and the Hubble distance
by the dashed curve.} 
\label{fig1}
\end{figure}
\begin{figure}[h]
\centerline{\includegraphics{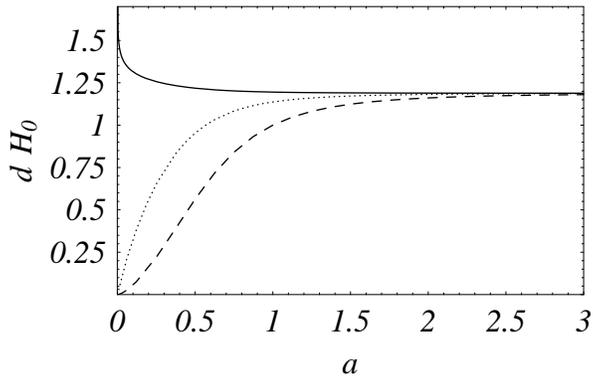}}
\caption{The same as in Fig.~1, but for $\epsilon=-0.1$.} 
\label{fig2}
\end{figure}

One may wonder why in this study of dark energy we insist on the
holographic point of view, when obviously the scaling law $G_N(a)$ and the
evolution law $a(t)$ as given by Eqs.~(5) (from which all phenomenological
implications
considered below emerge) are the same in any setup consisting of constant
$\rho_{\Lambda }$ and variable $G_N $. In addition, Eq.~(6) for the IR cutoff
appears as redundant here, since a dependence of $G_N $ on the scale factor
is
given directly by (5), with no obvious reference to holography. The
reason lies in the fact that it is nontrivially to provide a theoretical
background for the above setup sticking only with the Hilbert-Einstein
action and
variable cosmological parameters. In this context, scalar-tensor theories
appear
more natural for accommodating the above setup because there
generally $G_N \sim 1/\phi $, with no reference to $\rho_{\Lambda }$, which
can always be put in as a constant term in the action. In the present
situation, however,
one is forced to rely on RG approaches \cite{19, 20, 21} which employed the
Hilbert-Einstein action, and where both $G_N $ and $\rho_{\Lambda }$ varied
through the chosen evolving RG scale. From the point of view of our scenario
the problematic point in these approaches  is that the same mechanism is
responsible for the RG-running of both quantities. Therefore the RG approach
generally cannot support a scenario where only one quantity is varying.
On the other hand, one can show 
(see the first reference in \cite{7}) that the RG approach in quantum
gravity \cite{21} is manifestly underpinned by the generalized holographic
dark-energy relation (1), while the RG approach in a conventional
field-theoretical model in the classical curved background \cite{19, 20} can
also be supported by holography for certain choices of the RG scale
(see the second reference in \cite{7}).
Obviously, the generalized holographic principle (1) is able to accommodate
a larger class of models with running cosmological quantities, and for this
reason we find the model-independent holographic relation (1) a suitable
background for our setup. Although originally derived by Cohen et. al.
\cite{3} for the opposite limit, where $\rho_{\Lambda }$ is variable and 
$G_N $ is
static, we make use of the flexibility of the generalized relation (1) to
explore the opposite limit, where $\rho_{\Lambda }$ is static and $G_N $ is
variable. Therefore, the scale $\mu$ in (6) should not be confused with
any of the RG scales, and although with no practical meaning here, it serves
as an internal check for the holographic approach of Cohen et. al.. In
contrast to other approaches where the IR cutoff is  ad hoc chosen, it is
here unequivocally determined by dynamics, at the same time retaining its
intuitive
interpretation in measuring an 'extension' of the system (see Figs.~1 and 2).

Although $G_N \sim \mu^2 $ always in our model, we can notice from 
Figs.~1 and 2 a
different behavior of $\mu $ for small $a$ for each sign of $\epsilon $.
Specifically, it can be easily seen that for $a \rightarrow 0$,
$G_N \rightarrow 0$
for $\epsilon < 0$. Such a scale dependence implies that the coupling
$G_{N}$
is asymptotically free; a feature exhibited, for instance, by
higher-derivative
quantum gravity models at the 1-loop level \cite{julve}. Although
for positive sign
of $\epsilon $ there is no such  feature, in both cases the gravitational
coupling soon tends to a constant value, implying the absence of large
quantum gravity effects on cosmological scales.

In order to compare our model with observations, we need to adapt the
concept of effective EOS (for dark energy), as put forward by Linder and
Jenkins \cite{30}. 
%It is defined by the second term in the Hubble parameter
%$H$, which encapsulates the entire modification of the standard Hubble
%parameter,
It is defined by the second term in the Hubble parameter squared
that encapsulates the entire modification of the
standard Friedmann equation:
\begin{equation}
H^2=\frac{8\pi G_{N0}}{3}(\rho_{m0}a^{-3}+\rho_{\Lambda}^{eff}) ,
\end{equation}
as
\begin{equation}
w_{eff}(a) =-1-\frac{1}{3}\frac{a}{\rho_{\Lambda}^{eff}} 
\frac{d\rho_{\Lambda}^{eff}}{da} .
\end{equation}
For our model, $\rho_{\Lambda }^{eff}$ is given by
\begin{equation}
\rho_{\Lambda }^{eff}=\rho_{\Lambda }r_0 [-a^{-3}+A(a)], 
\end{equation}
so that
\begin{equation}
w_{eff}(a) =-1+\frac{-a^{-3}+A(a)B(a)}{-a^{-3}+A(a)} ,
\end{equation}
where
\begin{eqnarray}
& A(a)\equiv \displaystyle\frac{(r_0^{-1}+a^{-3+\epsilon})^{3/(3-\epsilon)}}
{(r_0^{-1}+1)^{\epsilon/(3-\epsilon)}} , & \nonumber \\
& B(a)\equiv \displaystyle\frac{3-\epsilon}{3} 
\frac{a^{-3+\epsilon}}{r_0^{-1}+a^{-3+\epsilon}} .
\end{eqnarray}
It is interesting to calculate the present-day value for $w_{eff}(a)$. We
obtain $w_{eff}(a = 1) = -1 -(r_0 /3)\epsilon $, giving $-0.986$ for 
$\epsilon =-0.1 $ and $r_0 = 3/7 $. Also, 
$dw_{eff}(a)/da|_{a=1} =\epsilon [(6-\epsilon)r_0^{-1}+3]/[2(1+r_0)]$,
giving $-0.40$
for $\epsilon = -0.1 $. One  sees that our $w_{eff}$ is maximally
elastic, in the sense that it can be made arbitrary close to $-1$, thus easily
complying with the recent data which give the EOS converging rapidly
towards $-1$. Hence, the `minimal' character of our dynamical dark-energy
model is evident. It is also interesting to examine the limit of the  vanishing
CC, i.e. when $r_0^{-1}=0$. 
Although the $\epsilon $-dependence is still present in the
scaling of $G_N$ and $\rho_m $ in this limit, their product becomes $\epsilon
$-independent, giving $\rho_{\Lambda }^{eff} = 0$, thus reducing
cosmology to the standard CDM case. In addition, one can be convinced 
that, asymptotically, our model
always gives a de Sitter universe for both signs of 
$\epsilon$, 
meaning that arguments leading to the Big Rip \cite{31} no longer apply 
here. The
reason for having $w_{eff} < -1$ for some time in the future 
lies in the modified expansion rate for matter caused by the variable $G_N$,
and not in the exotic nature of dark energy.    

However, 
the analysis of the recent data indicates a slightly
better fit for 
the time-varying EOS  than for a CC \cite{18}. Specifically, the   
EOS evolution from $> -1$ to
$< -1$ in the recent past is mildly favored for redshifts 
$z \sim 0.1 - 0.2$.\footnote{We mention here that there is a hint from more 
recent analyses \cite{nesseris,padmanabhan} that this is not so obvious.} 
However, to have
this property implemented in our scenario, we need to switch to curved
spaces. Namely, it is evident for  the above flat-space case that crossing
of the phantom line, $w_{eff} = -1$, occurs always in the (near) future (for
both signs of $\epsilon $). For instance, for $\epsilon = -0.1$, the phantom
line is crossed at $z \simeq -0.33$ $(a \simeq 1.5)$. To switch the crossing
of the 
phantom line from the near future to the recent past, 
we introduce another modification
of the standard Hubble parameter in the form of curvature, i.e. a term
$-k/a^2 $ in the Friedmann equation that modifies (9) in (7). 
This modifies the effective EOS to
\begin{equation}
w_{eff}(a) =-1+\frac{-a^{-3}+A(a)B(a)+2ba^{-2}/3}{-a^{-3}+A(a)+ba^{-2}} ,
\end{equation}
where
\begin{equation}
b\equiv\frac{8\pi}{3}\frac{r_0+1}{r_0}\frac{\Omega_{k0}}{1-\Omega_{k0}} ,
\end{equation}
and $\Omega_{k0}\equiv -k/H_0^2$.
In particular, for $r_0 = 3/7 $, $\epsilon = -0.1$,
 and $\Omega_{k0} = -0.0025$,
we get $w_{eff}  = -1$ for $z=0.124$, with $w_{eff}$ decreasing with $a$,
as it should. In addition, as today $w_{eff}$ behaves as
\begin{eqnarray}
& w_{eff}(1)=-1+\displaystyle\frac{2f -\epsilon}{3(r_0^{-1}+f)} , & 
\nonumber \\
& \left. \displaystyle\frac{d w_{eff}}{da} \right|_{a=1} =
\displaystyle\frac{ \epsilon (r_0^{-1}+b)
[(6-\epsilon)r_0^{-1}+3]/(r_0^{-1}+1) +2b(2b-\epsilon) }
{ 3(r_0^{-1}+b)^2 } , &
\end{eqnarray}
we see that the `minimal' character of the scenario is still preserved for
curved universes. Finally, in the limits $\rho_{\Lambda } \rightarrow 0$ or 
$a \rightarrow \infty $, our curved-space model behaves identically as the
flat-space one above, with no Big Rip occurrence in the future.   
Let us stress that crossing of the phantom line has recently been shown
\cite{sola_extra} to be 
a general feature of models with variable cosmological parameters. 

To this end, we investigate some thermodynamic features of the
present scenario and show that the $\epsilon $-parameter is restricted by the
GSL of gravitational thermodynamics to assume only
positive values. The positivity of $\epsilon$ entails creation of 
matter.
We start with the fact that in an ever accelerating universe
there {\it always} exists a future event horizon. Thus, analogously to the
black-hole horizon, it can be attributed some thermodynamical quantities, like
entropy and temperature. Although, in a strict sense, 
this has been proved for
a de Sitter horizon only \cite{33}, where the temperature is proportional to
the inverse apparent horizon, $\sim H $, 
many authors use to apply the same concept
when exploring the thermodynamical behavior of accelerated universes driven
by other sorts of dark energy \cite{34}. In these cases, when the degeneracy
between the apparent and the event horizon is broken, the horizon entropy
always refers to the entropy of the future event horizon, while the
not-well-defined temperature of the dark-energy fluid as well as the Hawking 
temperature are usually assumed to be the same as the de Sitter temperature
$\sim H$ (see, however, \cite{35}).

We seek for  information on the $\epsilon $-parameter by assuming that
the GSL of gravitational thermodynamics is satisfied. The GSL states that
the entropy of the event horizon plus the entropy of matter and radiation in
the volume within the horizon cannot decrease in time. The easiest way to gain
information on the $\epsilon $-parameter is by considering the change of
entropy in the asymptotic regime, $a \gg 1$. In this case, one should also
add the entropy of Hawking particles since it is conceivable that the CMB
temperature will drop below the Hawking temperature after some time in the
(distant) future (see the first reference in \cite{34}). 
Hence, we have\footnote{Since we are dealing
here with the `true' CC, the entropy of the dark-energy fluid 
inside the cosmological event horizon is equal to zero.}
\begin{eqnarray}
& \displaystyle\frac{d S_{tot}}{da}\geq 0 , & \nonumber \\
& S_{tot}=S_{e} +S_{m} +S_r + S_{Hawk} , &
\end{eqnarray}
where 
\begin{eqnarray}
& S_e=\displaystyle\frac{\pi d_e^2}{G_N}, & \nonumber \\
& S_m=\displaystyle\frac{\rho_m}{m}\frac{4\pi}{3}d_e^3 , & \nonumber \\
& S_r=\alpha T_{CMB}^3 \displaystyle\frac{4\pi}{3}d_e^3 , & \nonumber \\
& S_{Hawk} =\beta T_{Hawk}^3 \displaystyle\frac{4\pi}{3}d_e^3 , 
\end{eqnarray}
$d_e$ is the future event horizon, $T_{Hawk}=H/2\pi$ and
$\alpha $ and $\beta $ are order-of-one constants.
In the asymptotic 
regime, $dS/da $ for the particular entropies in (16) reads
\begin{eqnarray}
& \displaystyle\frac{dS_e}{da}\approx \pi H_0^{-2}G_{N0}^{-1}
 (1+r_0)^{(3+\epsilon)/(3-\epsilon)} r_0 2(6-\epsilon) a^{-4+\epsilon} , &
\nonumber \\
& \displaystyle\frac{dS_m}{da}\approx -\frac{\rho_{m0}}{m} H_0^{-3}
 (1+r_0)^{9/(2(3-\epsilon))} \frac{4\pi(3-\epsilon)}{3} a^{-4+\epsilon} , & 
\nonumber \\
& \displaystyle\frac{dS_r}{da}\approx -\alpha 4\pi T_{CMB0}^3 H_0^{-3}
 (1+r_0)^{9/(2(3-\epsilon))}  a^{-4}, & \nonumber \\
& \displaystyle\frac{dS_{Hawk}}{da}\approx \beta \frac{3r_0(3-\epsilon)}{4} 
 a^{-4+\epsilon} . &
\end{eqnarray}
We see that for $\epsilon < 0$ the left-hand side of (15) is dominated by
$dS_r /da $. Since the entropy of radiation inside the
horizon, $S_r $, always decreases with time, the GSL as given by (15)
cannot be satisfied for $\epsilon < 0$. On the other hand, 
for $\epsilon >0$,
the GSL can be satisfied provided the following constraint is obeyed:
\begin{equation}
(1+r_0)^{(3+\epsilon)/(3-\epsilon)} 2(6-\epsilon)
-\frac{H_0}{m}\frac{(1+r_0)^{(3+\epsilon)/(2(3-\epsilon))} (3-\epsilon)}
{2\pi} +\beta H_0^2 G_{N0} \frac{3(3-\epsilon)}{4\pi} \geq 0 .
\end{equation}
Notice that (18) entails a trivial bound on the mass of the CDM particles,
$m \gsim\, H_0$.

We would like to conclude with a few additional remarks and comments. One
notices that the 
positivity of the $\epsilon $-parameter, as predicted by the assumed
validity of the GSL, may, in a lesser extent,
pose a drawback on our model.
Namely, for $\epsilon > 0$, the crossing of the phantom line in the recent
past (for curved space) is always such that $w_{eff}$ increases with $a$,
a trend not confirmed by the data. Still, one should note that there
is only a marginal $(2 \sigma )$ evidence for the phantom-line crossing,
being, in addition, strongly  dependent on the subsampling of the SNe
dataset. On the other hand, although there appear strong  
arguments for believing in the validity of the GSL, one argues that in some
cases, comprising both phantom (see the third reference in \cite{34}) 
and nonphantom \cite{35} fluids,
the GSL might not be fulfilled. In addition, $\epsilon > 0$ corroborates 
Dirac's hypothesis. Hence, we see from the above discussion that arguments
towards either $\epsilon > 0$ or $\epsilon < 0$ can by no means be 
deemed as decisive.

Finally, we should stress that a more pressing issue is the `cosmic
coincidence problem', which, beyond anthropic considerations, has no
solution in the present scenario. In models  with the running CC and static
$G_N $, in which there is a continuous energy transfer between the CC and
matter (see the first reference in \cite{31}), 
the constant term in $\rho_{\Lambda }$ is crucial since otherwise a
transition between deceleration to acceleration cannot be obtained. 
On the other hand, holography
cannot underpin such models as, by Eq. (1), the constant term in
$\rho_{\Lambda }$ is always set to zero. Hence, although
such models can ameliorate the `cosmic coincidence problem' to
some extent, they do not comply with observation.  
It would be interesting to explore the generalized equation of continuity
within the holographic dark-energy model (1),
\begin{equation}
\dot{G}_{N}(\rho_{\Lambda } + \rho_m ) + G_N \dot{\rho }_{\Lambda } +
G_N (\dot{\rho }_{m} + 3H\rho_m )  = 0,
\end{equation}
in which both $\rho_{\Lambda }$ and $G_N $ are variable and the scaling 
of $\rho_m $ is different from $a^{-3}$.
In this way it would be possible
to see if such a scenario can keep
nice features of the present minimal model regarding observation, 
at the same time offering a solution to the `cosmic coincidence problem'. 
Any other
improvement of the minimal scenario, for instance, 
that with additional degrees of freedom in the form of scalar fields,
would certainly have much less predictive power.

{\bf Acknowledgments. } This work was supported by the Ministry of Science,
Education and Sport of the Republic of Croatia, and
partially supported through the Agreement between the Astrophysical
Sector, S.I.S.S.A., and the Particle Physics and Cosmology Group, RBI.

\end{document}